\documentstyle[preprint,aps,epsf]{revtex}
\tightenlines
\def\Journal#1#2#3#4{{#1} {\bf #2}, #3 (#4)}

\def\NPB{{\em Nucl. Phys.} B}

\def\AP{{\em Ann. Phys.}}

\def\PLB{{\em Phys. Lett.} B}
\def\PRL{\em Phys. Rev. Lett.}
\def\PR{{\em Phys. Rev.}}
\def\PRB{{\em Phys. Rev.} B}
\def\PRD{{\em Phys. Rev.} D}

\def\CMP{{\em Comm. Math. Phys.}}

\def\ZETF{{\em Zh. Eksp. Teor. Fiz.}}
\def\PRS{{\em Proc. Roy. Soc.}}
\def\JETP{{\em Sov. Phys. JETP}}

\newcommand{\be}  {\begin{equation}}
\newcommand{\ee}  {\end{equation}}
\newcommand{\bea} {\begin{eqnarray}}
\newcommand{\eea} {\end{eqnarray}}

\newcommand{\cK}  {{\cal K}}
\newcommand{\cD}  {{\cal D}}

\newcommand{\hf}  {{1\over2}}
\newcommand{\nonu}{\nonumber\\}
\newcommand{\vphi}{\varphi}
\newcommand{\tr}  {{\rm tr}}
\newcommand{\cL}  {{\cal L}}
\newcommand{\lm}  {\lim_{a \rightarrow 0}}
\newcommand{\br}  {\hskip -0.25cm /}
\newcommand{\cm}  {{\cal M}^2_L}
\newcommand{\ml}  {m_{LR}^2}
\newcommand{\dml} {\delta m_L^2}
\newcommand{\tml} {\tilde m_{LR}^2}
\newcommand{\tmr} {\tilde m_R^2}
\newcommand{\tmrp} {\tilde m_R^{2'}}
\newcommand{\mr}  {m_R^2}
\newcommand{\lr}  {\lambda_R}
\newcommand{\dl}  {\delta\lambda}
\newcommand{\adots} {{\mathinner{\mkern2mu\raise1pt\hbox{.}\mkern2mu
\raise4pt\hbox{.}\mkern2mu\raise7pt\hbox{.}\mkern1mu}}}

\begin{document}
\title{The antiferromagnetic $\phi^4$ Model, II. The one-loop renormalization}

\author{Vincenzo Branchina\thanks{branchina@crnvax.in2p3.fr}}
\address{Laboratory of Theoretical Physics, Louis Pasteur University\\
3 rue de l'Universit\'e 67087 Strasbourg, Cedex, France}

\author{Herv\`e Mohrbach\thanks{mohrbach@crnvax.in2p3.fr}}
\address{Laboratory of Theoretical Physics, Louis Pasteur University\\
3 rue de l'Universit\'e 67087 Strasbourg, Cedex, France\\
and\\
LPLI-Institut de Physique, F-57070 Metz, France}

\author{Janos Polonyi\thanks{polonyi@fresnel.u-strasbg.fr}}
\address{Laboratory of Theoretical Physics, Louis Pasteur University\\
3 rue de l'Universit\'e 67087 Strasbourg, Cedex, France\\
and\\
Department of Atomic Physics, L. E\"otv\"os University\\
Puskin u. 5-7 1088 Budapest, Hungary}
\date{\today}
\maketitle
\begin{abstract}
It is shown that the four dimensional antiferromagnetic lattice $\phi^4$ model 
has the usual non-asymptotically free scaling law in the UV regime around the chiral 
symmetrical critical point. The theory describes a scalar and a pseudoscalar particle.  
A continuum
effective theory is derived for low energies. A possibility of constructing a model
with a single chiral boson is mentioned.
\end{abstract}

\section{Introduction}
This is the second of two papers where we study the impact of higher derivative terms 
in field theories. In ref.\cite{bmp} we have pointed out that the presence of
these terms in a self interacting single component scalar field theory produces 
tree level effects which may drive the formation of new vacua which is not accounted 
for by the decoupling theorem \cite{apca}. Three phases have been studied in the $\Phi^4$  
theory by means of the mean-field approximation, the
paramagnetic, $<\Phi(x)>=0$, the ferromagnetic, $<\Phi(x)>=\Phi_1\not=0$ and the
antiferromagnetic where $<\Phi(x)>$ is an oscillating function. $2^d$ bands have
been found in the dispersion relation for the elementary excitations above these vacuua
in a certain range of the coupling constants of the d-dimensional theory. 
A reduced Brillouin zone was introduced for each band. Two zones 
describe particle like excitations and the others decouple in the mean-field continuum limit.

For a special choice of the coefficients of the higher order derivatives the theory possesses a 
formal chiral symmetry which allows us to decouple the two particles.
The two decoupled modes correspond to the sublattices consisting of the even and the odd 
lattice sites. The theory which has nonvanishing field variables on one of the sublattices
only is local and describes a chiral boson. In fact the space inversion exchanges
the two sublattices and there is no space inversion partner of the particle in such a model.

We extend the analysis \cite{bmp} in the upper critical dimension, $d=4$, 
for theories in the vicinity of the chiral invariant critical point to the one-loop order and show that 
the beta functions of the lattice regulated theory with our $O(\Box^2)$ term in the lagrangian 
are those of an ordinary $\Phi^4$ model and give a renormalized lagrangian in terms of 
the continuum field variables. The one-loop renormalizability turns out to be a nontrivial
consequence of the perturbative renormalizability around the critical point 
of the usual $\Phi^4$ model without higher order derivative terms because we have to
render the dynamics for the two particles finite by fine tuning the set of the parameters 
of the bare lagrangian with a single quantum field. 

There is a formal similarity between the tricritical point
of the $\phi^6$ model and the chiral symmetrical theory. The mean field
solution of the model with the potential
\be
V(\phi)={g_2\over2}\phi^2+{g_4\over4!}\phi^4+{g_6\over6!}\phi^6
\ee
shows a tricritical point at $g_4=0$ which separates the second and
the first order phase transition lines with different scaling laws \cite{chang}.
In our case the dispersion relation
\be
G^{-1}(p)=m^2+p^2-c_2a^2p^4+c_4a^4p^6
\ee
produces a tricritical point when $c_2$ is sufficiently large to
give an absolute minimum at nonvanishing values of the momentum.
When $G^{-1}$ at the minimum is negative an inhomogeneous
condensate is formed. The chiral symmetrical point where this happens
is the Lifshitz point. This was introduced in \cite{lif} where the 
$\epsilon$-expansion was used to find out the scaling laws.
A scalar model where the dispersion relation has a single minimum
at nonvanishing momentum was considered in \cite{frenkont}
and \cite{braz}. The phase transition towards the inhomogeneous
vacuum was identified in the mean field level and the quantum 
fluctuations were taken into account in \cite{braz}.
We will be working at $d=4$ and extend the loop computation into
the phase with inhomogeneous condensate which generates a "dangerous 
irrelevant variable" \cite{dang}. The dispersion relation of our model 
has several minima hence it contains several particle modes simultaneously.
The condensate formation mechanism selects one of these particle sectors 
in a manner reminiscent of the spontaneous symmetry breaking.

There is a technical problem to solve in achieving this goal because
more than one particle corresponds to the same quantum field. The
formal problem is that higher order derivative terms in the kinetic energy 
imply the presence of states with negative metric \cite{unit} and
may render the effective action complex. But we argue that far
below the momentum scale of the condensate we find only two particles,
both with positive metric. Their dispersion relations can be replaced in 
the continuum limit by the usual quadratic expressions coming from a 
manifestly hermitean free lagrangian \cite{riesz}. The situation
turns out to be somehow similar to the species doubling of the lattice fermions
where one finds several particle modes in the dispersion relation of a single
bispinor field. We introduce a $2^d$-component field variable,
$\Phi_\alpha(x)$, $\alpha=1,\cdots,2^d$, for the computation of the one-loop 
generator functional for the 1PI functions of the different excitation bands
and show that it can be made finite by an appropriate fine tuning of 
the coupling constants of the original lagrangian. 

The organization of the paper is the following. The basic tools of the perturbation 
expansion are developed in Section 2. The computation of the effective potential is
presented in Section 3. The elimination of the divergences is shown and the finite 
renormalized coupling constants are obtained in Section 4. Section 5 is devoted to
a simple effective theory which reproduces our model at low energy. A brief conclusion
is in Section 6.

\section{The perturbation expansion}
We develop the basic formula for the perturbation expansion in the
scalar $\phi^4$ model in $d=4$ with higher order derivatives 
by keeping the original field variable, $\Phi(x)$. As in \cite{bmp} the theory
is regularized on the lattice.

\subsection{The lagrangian}
The model considered contains a one component field variable, $\Phi(x)$, and is defined
by the bare, cut-off lagrangian,
\be
\cL=\hf\partial_\mu\Phi(x)\cK\biggl({(2\pi)^2\over\Lambda^2}\Box\biggr)\partial_\mu\Phi(x)
+{m^2_B\over2}\Phi^2(x)+{\lambda_B\over4}\Phi^4(x),\label{lagrc}
\ee
where 
\be
\cK(z)=1+c_2z.
\ee

We write this lagrangian as 
\be
\cL=\cL_1+\cL_2
\ee
with
\be
\cL_1=\hf\partial_\mu\Phi\cK\biggl({(2\pi)^2\over\Lambda^2}\Box\biggr)\partial_\mu\Phi
+{\mr\over2}\Phi^2+{\lr\over4}\Phi^4,
\ee
and
\be
\cL_2={\delta Z\over2}\partial_\mu\Phi\cK\biggl({(2\pi)^2\over\Lambda^2}\Box\biggr)\partial_\mu\Phi
+{\delta m^2\over2}\Phi^2 +{\dl\over4}\Phi^4.
\ee
We will use $\cL_1$ non-perturbatively in the selection of the saddle point
and $\cL_2$ perturbatively in removing the UV divergences of the loop-corrections.
The bare parameters are defined as $m^2_B=\mr+\delta m^2$ and $\lambda_B=\lr+\dl$. 
We have no counter terms for the coupling constants $c_j$ because their leading order 
renormalization is at the tree-level. 

We employ lattice regularization where one introduces the 
dimensionless variables $x^\mu$, $\vphi=a^{d/2-1}\Phi$, $m^2_L=m^2a^2$
and the unit vectors $(e_\mu)^\nu=\delta_{\mu\nu}$ and writes the action as
\be
S_1[\vphi]=\sum\limits_x\cL_1(x)=\sum\limits_x\biggl\{K[\vphi;x]+V(\vphi(x))\biggr\},
\ee
where
\bea
K[\vphi;x]&=&-\hf\vphi(x)\biggl[
A\vphi(x)+\sum_\mu\biggl(B(\vphi(x+e_\mu)+\vphi(x-e_\mu))\nonu
&+&C(\vphi(x+2e_\mu)+\vphi(x-2e_\mu)\biggr)\nonu
&+&\sum_{\mu\ne \nu}\biggl(E(\vphi(x+e_\mu+e_\nu)+2\vphi(x+e_\mu-e_\nu)
+\vphi(x-e_\mu-e_\nu)\biggr)\biggr]
\eea
and
\be
V(\vphi(x))={\ml\over2}\vphi^2(x)+{\lr\over4}\vphi^4(x),
\ee
The coefficients of the kinetic energy are
\bea
A&=&-2d+(4d^2+2d)c_2,\nonu
B&=&1-4dc_2,\nonu
C&=&c_2,\nonu
E&=&c_2.
\eea

The field variable, $\vphi=\vphi_{vac}+\phi$, is the sum of the tree-level vacuum,
\be
\vphi_{vac}(x)=\vphi_1+\vphi_{2^d}\chi(x),
\ee
where
\be
\chi(x)=(-1)^{\sum\limits_{\mu=1}^dx^\mu},
\ee
and the quantum fluctuations, $\phi(x)$. We will study the theory in the 
para- ($\vphi_1=\vphi_{2^d}=0$), ferro- ($\vphi_1\not=0,~~\vphi_{2^d}=0$) and the 
$(1,2)$ antiferromagnetic ($\vphi_1=0,~~\vphi_{2^d}\not=0$) phases in $d=4$.
The lagrangian for the quantum fluctuations is
\bea
\cL_{1P}&=&\hf\partial_\mu\phi(x)\cK\biggl({(2\pi)^2\over\Lambda^2}\Box\biggr)\partial_\mu\phi(x)
+{\ml\over2}\phi^2(x)+{\lr\over4}\phi^4(x),\nonu
\cL_{1F}&=&\hf\partial_\mu\phi(x)\cK\biggl({(2\pi)^2\over\Lambda^2}\Box\biggr)\partial_\mu\phi(x)
+{1\over2}(\ml+3\lr\vphi_1^2)\phi^2(x)\nonu
&&+\lr\vphi_1\phi^3(x)+{\lr\over4}\phi^4(x),\nonu
\cL_{1AF}&=&\hf\partial_\mu\phi(x)\cK\biggl({(2\pi)^2\over\Lambda^2}\Box\biggr)\partial_\mu\phi(x)
+{1\over2}(\ml+3\lr\vphi_{2^d}^2)\phi^2(x)\nonu
&&+\lr\vphi_{2^d}\chi(x)\phi^3(x)+{\lr\over4}\phi^4(x),
\eea
\bea
\cL_{2P}&=&{\delta Z\over2}\partial_\mu\phi(x)\cK\biggl({(2\pi)^2\over\Lambda^2}\Box\biggr)\partial_\mu\phi(x)
+{\dml\over2}\phi^2(x)+{\dl\over4}\phi^4(x),\nonu
\cL_{2F}&=&{\delta Z\over2}\partial_\mu\phi(x)\cK\biggl({(2\pi)^2\over\Lambda^2}\Box\biggr)\partial_\mu\phi(x)
+\vphi_1(\dml+\dl\vphi_1^2)\phi(x)\nonu
&&{1\over2}(\dml+3\dl\vphi_1^2)\phi^2(x)+\dl\vphi_1\phi^3(x)
+{\dl\over4}\phi^4(x),\nonu
\cL_{2AF}&=&{\delta Z\over2}\partial_\mu\phi(x)\cK\biggl({(2\pi)^2\over\Lambda^2}\Box\biggr)\partial_\mu\phi(x)
+\vphi_{2^d}\chi(x)(\dml+\dl\vphi_{2^d}^2)\phi(x)\nonu
&&{1\over2}(\dml+3\dl\vphi_{2^d}^2)\phi^2(x)
+\dl\vphi_{2^d}\chi(x)\phi^3(x)+{\dl\over4}\phi^4(x),
\eea
where the tree-level vacuum is given by
\bea
P:&\vphi_{P1}=0&\vphi_{P2^d}=0,\nonu
F:&\vphi_{F1}=-{\ml\over\lambda}&\vphi_{F2^d}=0,\nonu
AF:&\vphi_{AF1}=0&\vphi_{AF2^d}=-{\ml+\cm\over\lambda}.
\eea
Here $\cm$ stands for the eigenvalue of the the kinetic energy 
on the antiferromagnetic vacuum, 
\be
{(2\pi)^2\over\Lambda^2}\Box\cK\biggl({(2\pi)^2\over\Lambda^2}\Box\biggr)\chi=
\cm(d,c_2)\chi,
\ee
with
\be
\cm(d,c_2)=4d\cK(-4d)=4d(1-4dc_2).
\ee
The tree-level conditions for the three phases shown in Fig. 1 are
\bea
P:&\ml\ge0&\ml+\cm\ge0,\nonu
F:&\ml\le0&\cm\ge0,\nonu
AF:&\ml+\cm\le0&\cm\le0.\nonu
\eea

\subsection{The free propagator}
The free propagator,
\be
<\phi(x)\phi(y)>=\int_{p\le\pi}{d^dp\over(2\pi)^d}e^{-ipx}G(p),
\ee
is given by
\be
G^{-1}(p)=\tml+\hat p_\mu\hat p^\mu\cK(-\hat p_\mu\hat p^\mu),
\ee
where mass parameter with the tilde includes the shift due to the condensate
\be
\tml=\cases{\ml&P,\cr-2\ml&F,\cr-2\ml-3\cm(d,c_2)&AF,}\label{shmass}
\ee
in the different phases and
\be
\hat p_\mu=2\sin{p_\mu\over2}.
\ee
We further write
\be
G^{-1}(p)={\cal P}^2(p)-c_2{\cal P}^4(p)+\tml,
\ee
with the help of 
\be
{\cal P}^2(p)=4\sum_\mu\sin^2{p^\mu\over2}.
\ee

It is advantageous to divide the Brillouin zone,
\be
{\cal B}=\bigg\{k_\mu,~|k_\mu|\le\pi\biggr\},
\ee
into $2^d$ restricted zones,
\be
{\cal B}_\alpha=\biggl\{|k_\mu-P_\mu(\alpha)|\le{\pi\over2}\biggr\},
\ee
whose centers are at
\be
P_\mu(\alpha)=\pi n_\mu(\alpha),
\ee
where $n_\mu(\alpha)=0,1$ and the index $1\le\alpha\le2^d$ is given by 
\be
\alpha=1+\sum_{\mu=1}^d\alpha_\mu 2^{\mu-1}.
\ee
The propagator for the zone ${\cal B}_\alpha$ is
\be
G_\alpha(p)=G(P(\alpha)+p).
\ee
It turns out that only the Brillouin zones $\alpha=1$ and $2^d$ contain particle like
excitations and the corresponding propagators are
\be
G^{-1}_\alpha(p)=\tml(\alpha)+Z(\alpha)p^2+O(p^4),
\ee
where the mass and the wave function renormalization constant are given in Table 1.
Note that $\tml(1)=\tml$.

The fact that the vacuum is a single Fourier mode 
offers the possibility of recovering the energy-momentum conservations
in the anti-ferromagnetic phase. The possible translations which keep the
vacuum invariant consist of an even number of shift of the integer lattice
coordinates. The corresponding spectrum of the momentum operator is
\be
p_{AF\mu}=p_\mu~({\rm mod}\pi).\label{consm}
\ee
In fact, the function ${\rm mod}\pi$ substracts the part of the momentum which 
can be exchanged with the antiferromagnetic vacuum and the resulting value is conserved.
In this manner the momentum non-conservation is traded for the exchange of the
particle type, the "flavor dynamics".

\subsection{Chiral symmetry}
The chiral transformation
\be
\chi:~~~~~~~~\phi(x)\longrightarrow\chi(x)\phi(x),\label{chtrrs}
\ee
which appears as the shift
\be
p_\mu\to p_\mu+P_\mu(2^d)
\ee
in the Fourier space is a symmetry of the lagrangian when
\be
c_2={1\over4d},~~~~c_4=0.\label{chinth}
\ee
The two particle species are degenerate in the chiral invariant theory.

The operator ${\cal P}_\pm=\hf(1\pm\chi)$ identifies the fields which belong to the 
even or odd sublattices, 
\be
{\cal P}_\pm\phi_\pm=\phi_\pm.\label{slproj}
\ee
The chiral transformation is represented by
\be
\phi_\pm\to\pm\phi_\pm,
\ee
so the chiral fields $\phi_+$ and $\phi_-$ decouple in the chiral invariant theory.
The inversion of odd number of coordinates exchanges the chiral fields. The low energy
excitations in ${\cal B}_1$ and ${\cal B}_{16}$ correspond to 
\be
\tilde\phi_\pm=\phi_+\pm\phi_-,
\ee
where the fields $\phi_\pm$ are slowly varying.
Thus the low energy excitations of the zones ${\cal B}_1$ and ${\cal B}_{16}$ 
have space inversion parity $+1$ and $-1$, respectively.

\section{The Effective Potential}
The renormalization of the theory will be performed in the para- ferro and the
$(1,2)$ antiferromagnetic phase in the one-loop order by making the 
effective potential cut-off independent. It is easy to verify that this latter
is enough, i.e. the wavefunction renormalization constant is finite at the one-loop
order, $\delta Z=0$.

\subsection{A one-loop diagram}
In order to develop the appropriate notation we consider first a simple example,
the contribution of the second graph of Fig. 2 in the most complicated case, 
the $(1,2)$ antiferromagnetic phase,
\be
\Sigma(k)=\hf\lambda^2_R\vphi_{2^d}^2\int_{p\le\pi}{d^dp\over(2\pi)^d}
G(k+p)G(p+P(2^d))\label{sigm}.
\ee
The lattice cut-off of the loop integrals, $p<\pi$, should always be understood
as the constraint $|p_\mu|<\pi$, for $\mu=1,\cdots,d$ imposed on the torus ${\cal D}$
unless it is stated otherwise.
An integral like this can be written in a simpler form by the help of the
following matrix notation. The loop integration is split into the sum over
the $2^d$ restricted Brillouin zones,
\be
\int_{p\le\pi}d^dpf(p)=\sum_{\alpha=1}^{2^d}\int_{p\le\pi/2}d^dpf(P(\alpha)+p),
\ee
in particular,
\be
\int_{p\le\pi}d^dpG(p)=\sum_{\alpha=1}^{2^d}\int_{p\le\pi/2}d^dp
f(P(\alpha)+p)=\sum_{\alpha=1}^{2^d}\int_{p\le\pi/2}d^dpG_\alpha(p).
\ee
Returning to our one-loop integral (\ref{sigm}) we find
\bea
\Sigma_c(k)=\hf\lr^2\vphi_{2^d}^2\sum_{\alpha=1}^{2^d}
\int_{p\le\pi/2}{d^dp\over(2\pi)^d}G_\alpha(k+p)G_{\bar\alpha}(p),
\eea
where we have introduced the region complementer to $\alpha$,
\be
\bar\alpha=2^d+1-\alpha.
\ee

To simplify further the latter expression we now promote $\alpha$
to be an internal index distinguishing different kind of fluctuations
and define the propagator,
\be
G_{\alpha,\beta}(p)=\delta_{\alpha,\beta}G(P(\alpha)+p.),\label{prop}
\ee
which is diagonal in this new internal space. The contribution considered to the
self energy is then written in matrix notation,
\be
\Sigma(k)_c=\hf\lr^2\vphi_{2^d}^2\int_{p\le\pi/2}{d^dp\over(2\pi)^d}
\tr\bigl[G(k+p)\gamma^{2^d}G(p)\gamma^{2^d}\bigr],\label{gnull}
\ee
by the help of the matrix
\be
\gamma^{2^d}_{\alpha,\beta}=\delta_{\alpha+\beta,2^d+1},
\ee
which describes the change of the type of particle after scattering off the vacuum.

\subsection{The one-loop effective potential}
Let us denote the usual 1PI functions by $\Gamma^{(n)}(p_1,\cdots,p_n)$.
The 1PI function for the excitations of the type $\alpha_1,\cdots,\alpha_n$
is given as 
\be
\Gamma^{(n)}(P(\alpha_1)+p_1,\cdots,P(\alpha_n)+p_n).
\ee
The generator function for the zero momentum excitations, the effective potential, is 
defined as 
\be
V_{eff}(\Phi)=\sum_{n=0}^\infty{1\over n!}
\sum_{\alpha_1,\cdots,\alpha_n}\Phi_{\alpha_1}\cdots
\Phi_{\alpha_n}\Gamma^{(n)}(P(\alpha_1),\cdots,P(\alpha_n)).
\ee

The matrix $\gamma^{2^d}$ in (\ref{gnull}) reflects a modification of the Feynman rules.
Whenever a propagator $G_\alpha(p)$ is inserted in a graph it contains the momentum 
$P_\mu(\alpha)+p$. We keep track of the first term of this sum by introducing
a $2^d$-component field, $\Phi_\alpha$, in such a manner that the 
$\alpha$-th component will be responsible of the excitations in
${\cal B}_\alpha$. Thus the Feynman rules are
those of a $2^d$-component field with the propagator (\ref{prop}) and each 
external line with $p=0$ is represented by the insertion of the matrix 
\be
\Phi\br=\sum_{\alpha=1}^{2^d}\gamma^\alpha\Phi_\alpha,
\ee
where
\be
\gamma^\alpha_{\rho,\sigma}=\prod_{\mu=1}^d
\delta_{\sigma_\mu+\alpha_\mu-\rho_\mu(mod2),0}
\ee
takes care of the change of the particle type at each vertex due to the
momentum flowing from the external leg. We will use either the index
$\alpha$ or its vector representative, $n_\mu(\alpha)$, in the formulae.

Taking advantage of the matrix formalism introduced above we obtain
\bea
V_{eff}(\Phi)&=&\hf\int_{p\le\pi/2}{d^dp\over(2\pi)^d}
\tr\ln[{\cal P}^2(P+p)\cK(-{\cal P}^2(P+p))\nonu
&&+\tml+6\lr\Phi\br\vphi\br+3\lr\Phi\br^2],\nonu
&=&\hf\int_{p\le\pi/2}{d^dp\over(2\pi)^d}
\tr\ln[{\cal P}^2(P+p)\cK(-{\cal P}^2(P+p))\nonu
&&+\tml+3\lr(\Phi\br+\vphi\br)^2],
\eea
where the matrix $P$ is given by
\be
P_{\alpha,\beta}=\delta_{\alpha,\beta}P(\alpha)
\ee
and the vacuum field is
\be
\vphi\br=\vphi_1\gamma^1+\vphi_{2^d}\gamma^{2^d}.
\ee 

The complete one-loop effective potential 
$V^{(0)}(\Phi)+V^{(1)}_{eff}(\Phi)$ for the background field 
\be
\Phi\br=\Phi_1\gamma^1+\Phi_{2^d}\gamma^{2^d},
\ee
is obtained in (\ref{varba}) and (\ref{effpoti}),
\bea 
V^{P(0)}(\Phi)&=&\hf\biggl({\cal P}^2(P(1))\cK(-{\cal P}^2(P(1)))+\ml+\dml\biggr)\Phi_1^2\nonu
&&+\hf\biggl({\cal P}^2(P(2^d))\cK(-{\cal P}^2(P(2^d)))+\ml+\dml\biggr)\Phi_{2^d}^2\nonu
&&+{\lambda+\dl\over4}(\Phi_1^4+\Phi_{2^d}^4 +6\Phi_1^2\Phi_{2^d}^2),\nonu
V^{F(0)}(\Phi)&=&V^{P(0)}(\Phi+\vphi_F),\nonu
V^{AF(0)}(\Phi)&=&V^{P(0)}(\Phi+\vphi_{AF}),\nonu\label{treepot}
\eea
and
\bea
V^{P(1)}_{eff}(\Phi)&=&\hf\int_{p\le\pi/2}{d^dp\over(2\pi)^d}
\sum_{\alpha=1}^{2^{d-1}}\ln\biggl\{\nonu
&&\times\biggl[{\cal P}^2(P(\alpha)+p)\cK(-{\cal P}^2(P(\alpha)+p))\nonu
&&+\ml+3\lr(\Phi_{2^d}^2+\Phi^2_1)\biggr]\nonu
&&\biggl[{\cal P}^2(P(\bar\alpha)+p)\cK(-{\cal P}^2(P(\bar\alpha)+p))\nonu
&&+\ml+3\lr(\Phi_{2^d}^2+\Phi^2_1)\biggr]
-36\lambda^2_R\Phi_1^2\Phi_{2^d}^2\biggr\}\nonu
V^{F(1)}_{eff}(\Phi)&=&V^{P(1)}(\Phi+\vphi_F),\nonu
V^{AF(1)}_{eff}(\Phi)&=&V^{P(1)}(\Phi+\vphi_{AF}) \label{loopot}.
\eea
The mass parameter of the effective potential in the ferro- and the antiferromagnetic
phase after the shift $\Phi\to\Phi+\vphi$ is given by (\ref{shmass}).

\section{The renormalization in $d=4$}
The divergences arising in the one-loop integral for the effective
potential are isolated by expanding the logarithm in the integrand.
We reintroduce the lattice spacing and use dimensional quantities in the 
rest of this paper. One finds three divergent integrals,
\bea
\cD_1&=&\sum_{\alpha=1}^{16}\int_{p\le{\pi\over2a}}{d^4p\over(2\pi)^4}G_\alpha(p)\nonu
\cD_2&=&\sum_{\alpha=1}^{16}\int_{p\le{\pi\over2a}}{d^dp\over(2\pi)^4}G_\alpha(p)^2\nonu
\bar\cD_2&=&\sum_{\alpha=1}^8\int_{p\le{\pi\over2a}}{d^4p\over(2\pi)^4}
G_\alpha(p)G_{\bar\alpha}(p),
\eea
and the divergent part of $V^{AF(1)}(\Phi)$ turns out to be
\bea
\label{effpdiv}
V^{AF(1)}_{div}(\Phi)&=&{C\over2}\cD_1-{C^2\over4}\cD_2-{B^2\over2}\bar\cD_2\nonu
&=&{3\over2}\lr\cD_1[(\Phi_{16}+\vphi_{16})^2+\Phi_1^2]\\
&&-{1\over8}\lr^236\bar\cD_2[(\Phi_{16}+\vphi_{16})^4+\Phi_1^4
+6(\Phi_{16}+\vphi_{16})^2\Phi_1^2]\nonu
&&-{9\over4}\lr^2\Delta\cD_2\biggl[(\Phi_{16}+\vphi_{16})^4+\Phi_1^4
-{4\over3}(\Phi_{16}+\vphi_{16})^2\Phi_1^2\biggr],\nonumber
\eea
where
\be
\Delta\cD_2=\cD_2-2\bar\cD_2,
\ee
and
\bea
B&=&6\lr\Phi_1(\phi_{16}+\Phi_{16}),\nonu
C&=&3\lr(2\vphi_{16}\Phi_{16}+\Phi_{16}^2+\Phi^2_1).
\eea
The corresponding expression for the ferromagnetic phase can be obtained by
the exchange $1\leftrightarrow16$ of the internal index. The condensate
has to be set to zero, $\vphi_\alpha=0$, in the expressions of the paramagnetic phase.

The choice of the mass counterterm,
\be
\dml=-3\lr\cD_1-9\lr^2\vphi_{16}^2\cD_2\label{ctmm},
\ee
is straightforward after comparing (\ref{effpdiv}) with (\ref{treepot}).
But there is a problem with the counterterm $\dl$ because it can not
eliminate the divergences for both particles in the same time when 
$\Delta\cD_2\not=0$.

\subsection{Renormalization with chiral symmetry}
The remedy of the problem of the divergences $O(\Phi^4)$ comes from the
observation that the chiral symmetry protects against the unwanted
divergences. In fact, the chiral transformation, (\ref{chtrrs}), acts as
\bea
\Phi_\alpha&\to&\Phi_{\bar\alpha},\nonu
G_\alpha(p)&\to&G_{\bar\alpha}(p)
\eea
on the variables of the effective potential and the propagator and the chiral
symmetry requires
\be
G_{\alpha}(p)=G_{\bar\alpha}(p)\label{symtr}
\ee
which reduces the number of divergences since we gain the relation
\be
\Delta\cD_2=\cD_2-2\bar\cD_2=0.
\ee
The divergent part of the effective potential is now written as
\bea
V^{AF(1)}_{div}(\Phi)&=&3\lr\cD_1[(\Phi_{16}+\vphi_{16})^2+\Phi_1^2]\nonu
&-&9\lr^2\bar\cD_2[(\Phi_{16}+\vphi_{16})^4+\Phi_1^4+
6(\Phi_{16}+\vphi_{16})^2\Phi_1^2]. 
\eea
Comparing it with (\ref{treepot}) we arrive at the choice
\be
\dl=18\lr\bar\cD_2\label{ctml}.
\ee

Thus one can eliminate the divergences of the chiral symmetrical 
theory in either of the phases by the help of
the appropriate fine tuning of the parameters $m^2_B$ and $\lambda_B$
of the original lagrangian. The chiral invariant theory is invariant
under the exchange of the two degenerate particles. Using the chiral fields,
$\Phi_\pm=\Phi_1\pm\Phi_{16}$,
one can decouple the two particle modes. Let's consider for example the case of 
the paramagnetic phase. Replacing in (\ref{treepot}) the appropriate values of 
$\dml$ and $\dl$, as given respectively in (\ref{ctmm}) and (\ref{ctml}), from 
(\ref{treepot}) 
and (\ref{loopot}) we get for the effective potential along the chiral line $\chi_P$ 
(see Fig.1),  
\be
V^P_{eff}(\Phi_1,\Phi_{16})=V_{eff}^{ch}(\Phi_+)+V_{eff}^{ch}(\Phi_-),
\ee
where
\bea
V_{eff}^{ch}(\Psi)&=&\hf \mr\Psi^2+{1\over4}\lr\Psi^4
+\hf\sum\limits_{\alpha=1}^{2^{d-1}}\int_{p\le{\pi\over2a}}
{d^4p\over(2\pi)^4}\nonu
&&\times\ln\biggl[(P(\alpha)+p)^2
\cK\biggl(-{(2\pi)^2\over\Lambda^2}(P(\alpha)+p)^2\biggr)
+\mr+6\lr\Psi^2\biggr]\nonu
&&-3\lr\Psi^2\int_{p\le{\pi\over2a}}{d^4p\over(2\pi)^4}{1\over p^2+\mr}\\
&&+9\lr^2\Psi^2\int_{p\le{\pi\over2a}}{d^4p\over(2\pi)^4}{1\over(p^2+\mr)^2}.
\nonumber
\eea
The same is true along the chiral lines in the other phases. This decoupling 
arises because in either of the phases at the chiral line the lattice decouples 
into two different sublattices as explained in Ref. \cite{bmp}.

\subsection{Renormalization around the symmetrical point}
In order to remove the symmetry with respect to the exchange of the two particles
we consider the four dimensional theory with the tree-level cut-off dependence
\be
\ml=\mr a^2,~~~c_2={1\over16}\biggl[1+\sigma(a\mu)^{2+\kappa}\biggr],
\label{rtraj}
\ee
where $\sigma=\pm1$ and $\mu$ is a mass parameter to characterize the split
of the degeneracy in the spectrum,
\be
\cm=-16\sigma(a\mu)^{2+\kappa}.
\ee
The quantities referring to the symmetrical 
theory, $\mu=0$, will be labelled with a star. Since there
are several possibilities in reaching the continuum limits as shown in 
Fig.1b we collect the corresponding conditions for (\ref{rtraj}) in Table 2.
We will find that the one-loop corrections do not change qualitatively the tree-level 
spectrum. The degeneracy, $\mr(2^d)=\mr(1)$, is achieved analytically at the chiral line 
indicating that the chiral symmetry is not broken dynamically. We find $\mr(1)<\mr(2^d)$
in the phase $F$ and in the region $P_F$ ($P_F$ is the region of the paramagnetic phase 
on the left side of the chiral line $\chi_P$, see Fig.1b). On the contrary, $\mr(2^d)<\mr(1)$ 
in the regions $AF$ and $P_{AF}$ (on the right side of the chiral line $\chi_P$ in Fig.1b). 
We found no singularities in the effective potential due to the discontinuity 
in the momentum of the condensate when the chiral line is reached from the phases
$F$ or $AF$.

The complication we face is that there will be finite $\mu$-dependent
corrections from the counterterms in the vicinity of the
symmetrical theory, $\cD_2\not=2\bar\cD_2$. 
The detailed study of the $\mu$ dependence in the limit
$a\to0$ is presented in Appendix \ref{mudep}. One finds that the 
$\mu$-dependence drops out from the finite part of the effective
potential and is finite for $\cD_2$ and $\bar\cD_2$. By introducing
\bea
\cD_2&=&\cD_2^\star+\delta\cD_2,\nonu
\bar\cD_2&=&\bar\cD^\star_2+\delta\bar\cD_2,
\eea
with $\cD_2^\star=2\bar\cD_2^\star$ one finds the finite expressions
\bea
\delta\cD_2&=&-{1\over16\pi^2}\ln{\tmr(16)\over\tmr(1)},\nonu
\delta\bar\cD_2&=&-{1\over16\pi^2}{\tmr(1)\over\tmr(16)-\tmr(1)}\ln{\tmr(16)\over\tmr(1)},
\eea
with 
\bea
\tmr(1)=\cases{\mr&P\cr-2\mr&F\cr-2\mr+48\mu^2(a\mu)^\kappa&AF,}
\eea
and
\bea
\tmr(16)=\cases{\mr-16\sigma\mu^2(a\mu)^\kappa&P,\cr-2\mr+16\mu^2(a\mu)^\kappa&F,\cr
-2\mr+32\mu^2(a\mu)^\kappa&AF.}
\eea
These expressions lead to the counterterms
\bea
\delta m^2&=&-3\lr\cD_1-9\lr^2\vphi_{16}^2\cD_2,\nonu
\delta \lambda&=&18\lr^2\bar\cD_2^\star.\label{countrt}
\eea

It is well known that the spontaneous symmetry breaking in a ferromagnetic
theory changes the counterterms by a cut-off independent finite piece
and influences the renormalization group flow at finite energies only. One
could, in principle, encounter a different situation in the antiferromagnetic phase
because the condensate is formed at the cut-off scale. 
Furthermore one band of the elementary 
excitations, in ${\cal B}_{16}$, belongs to the staggered modes which show fast oscillation
at the cut-off scale. It is the fine tuning of the value of the minimum of the dispersion
relation in the zone ${\cal B}_{16}$ which eliminates the divergent phase dependence
in the counterterms and restricts the effects of the phase transitions 
in the infrared region.

\subsection{Mass spectrum}
We are now in the position to follow the renormalization in the vicinity
of the critical system. The effective potential is written as the sum 
of the finite and divergent part,
\be
V^{(1)}_{eff}=V^{(1)}_{fin}+V^{(1)}_{div},
\ee
where the second term in the right hand side is defined by (\ref{effpdiv}).
One should bear in mind that starting with a single mass parameter in the bare
lagrangian we have already introduced different masses for the propagators
in the zones ${\cal B}_1$ and ${\cal B}_{16}$. The physical masses which contain the 
radiative corrections are given by the derivative of the effective potential.  

\underline{The Brillouin zone ${\cal B}_1$:} The mass square of the excitations is given by
\bea
\partial^2_{\Phi_1}V^{AF}_{eff}(\Phi)\Big\vert_{\Phi=0}
&=&\tmr+\delta m^2+3(\lr+\dl)\vphi_{16}^2\nonu
&&+3\lr\cD_1-9\lr^2\vphi_{16}^2\cD_2\nonu
&&-36\lr^2\vphi_{16}^2\bar\cD_2 
+\partial^2_{\Phi_1}V_{fin}^{AF(1)}(\Phi)\Big\vert_{\Phi=0}\nonu
&=&m_{ph}^2(1)
\eea
By the help of the counterterms (\ref{countrt}) we find
\be
m^2_{ph}(1)=\tmr+3\lr\vphi_{16}^2(1-3\lr\delta\cD_2-12\lr\delta\bar\cD_2)
+\partial^2_{\Phi_1}V_{fin}^{AF(1)}(\Phi)\Big\vert_{\Phi=0},
\ee
where
\bea
\delta\cD_2&=&-{1\over16\pi^2}\ln{-\tmr+16\mu^2(a\mu)^\kappa
\over-\tmr+24\mu^2(a\mu)^\kappa},\\
\delta\bar\cD_2&=&{1\over16\pi^2}\biggl(1+{-\tmr+16\mu^2(a\mu)^\kappa
\over-8\sigma\mu^2(a\mu)^\kappa}
\ln{-\tmr+16\mu^2(a\mu)^\kappa\over-\tmr+24\mu^2(a\mu)^\kappa}\biggr).\nonu
\eea
The computation of the finite part of the effective potential in Appendix C yields
vanishing result for the second derivatives with respect either field variables in all
phases. So we arrive at
\be
m^2_{ph}(1)=-2\mr+48\mu^2(\kappa)-18\lr^2\vphi_{16}^2\Delta\cD_2(\kappa),
\ee
in the continuum limit with
\be
\mu^2(\kappa)=\lm\mu^2(a\mu)^\kappa=\cases{0&$\kappa>0$,\cr\mu^2&$\kappa=0$,\cr
\infty&$\kappa<0$,}
\ee
and
\bea
\Delta\cD_2(\kappa)&=&\lm[\cD_2(\mu)-2\bar\cD_2(\mu)]\nonu
&=&\lm[\delta\cD_2-2\delta\bar\cD_2]\\
&=&\cases{{\tmr(1)\over\tmr(1)-\tmr(16)}\ln{\tmr(16)\over\tmr(1)}&$\kappa=0$,
\cr0&$\kappa>0$.}\nonumber
\eea

\underline{The Brillouin zone ${\cal B}_{16}$:} One finds
\bea
\partial^2_{\Phi_{16}}V^{AF}_{eff}(\Phi)\Big\vert_{\Phi=0}
&=&G^{-1}_{16}(0)+\delta m^2+3\dl\vphi_{16}^2+3\lr\cD_1\nonu
&&-36\lr^2\vphi_{16}^2\cD_2
+\partial^2_{\Phi_{16}}V_{fin}^{AF(1)}(\Phi)\Big\vert_{\Phi=0}\nonu
&=&-2\tmr+32\mu^2(a\mu)^\kappa+27\lr^2\vphi_{16}^2(2\delta\bar\cD_2
-\delta\cD_2)\nonu
&=&m_{ph}^2(16),
\eea
which results
\be
m_{ph}^2(16)=-2\tmr+32\mu^2(\kappa)-27\lr^2\vphi_{16}^2\Delta\bar\cD_2(\kappa).
\ee

Our conclusion is that for $\kappa>0$ $\delta\bar\cD_2=\delta\cD_2=0$ 
so the two particles become degenerate and the chiral symmetry is restored
in the continuum limit. For $\kappa=0$ the mass spectrum
stays non-degenerate. Finally the masses diverge as expected when $\kappa<0$.

\underline{The ferromagnetic phase:} We have, in a similar manner
\bea
m^2_{ph}(1)&=&-2\mr-18\lr^2\vphi_1^2\Delta\cD_2(\kappa)\\
m_{ph}^2(16)&=&-2\tmr+16\mu^2(\kappa)-27\lr^2\vphi_1^2\Delta\bar\cD_2(\kappa).\nonumber
\eea

\underline{The paramagnetic phase:} The renormalized masses for the line $\chi_P$ are 
\bea
m^2_{ph}(1)&=&\mr,\nonu
m_{ph}^2(16)&=&\mr,
\eea
showing the presence of the chiral symmetry. In the remaining part of the
paramagnetic phase we find a non-degenerate spectrum,
\bea
m^2_{ph}(1)&=&\mr,\nonu
m_{ph}^2(16)&=&\mr+16\mu^2(\kappa).
\eea
The $P-AF$ transition line corresponds to the spectrum
\bea
m^2_{ph}(1)&=&\mr,\nonu
m_{ph}^2(16)&=&16\mu^2\lm(\mu a)^\kappa.
\eea

\subsection{Coupling constant renormalization}
\underline{The Brillouin zone ${\cal B}_1:$} The definition of the renormalized coupling 
constant is
\bea
\partial^4_{\Phi_1}V^{AF}_{eff}(\Phi)\Big\vert_{\Phi=0}
&=&6(\lr+\dl)-54\lr^2\cD_2+\partial^4_{\Phi_1}V_{fin}^{AF(1)}(\Phi)
\Big\vert_{\Phi=0}\nonu
&=&6\lambda_{ph}(1),
\eea
giving
\bea
\lambda_{ph}(1)=(\lr+\dl)-9\lr^2(\cD^\star_2+\delta\cD_2)
+{1\over6}\partial^4_{\Phi_1}V_{fin}^{AF(1)}(\Phi)\Big\vert_{\Phi=0}.
\eea
With our choice of the counterterms we have
\be
\lambda_{ph}(1)=\lr-9\lr^2\lm\delta\cD_2+{1\over 6}\lm\partial^4_{\Phi_1}
V_{fin}^{AF(1)}(\Phi)\Big\vert_{\Phi=0}
\ee
in the continuum limit. 

\underline{The Brillouin zone ${\cal B}_{16}:$} The self-coupling constant for the field
$\Phi_{16}$ is
\be
\lambda_{ph}(16)=\lr-9\lr^2\lm\delta\cD_2
+{1\over 6}\lm\partial^4_{\Phi_{16}}V_{fin}^{AF(1)}(\Phi)\Big\vert_{\Phi=0}.
\ee

For the coupling constant which mixes the two fields we have
\bea
\partial^2_{\Phi_1}\partial^2_{\Phi_{16}}
V_{eff}^{AF}(\Phi)\Big\vert_{\Phi=0}&=&6(\lr+\dl)-18\lr^2(\cD^\star_2+\delta\cD_2)\nonu
&&-72\lr^2(\bar\cD_2^\star+\delta\bar\cD_2)
+\partial^2_{\Phi_1}\partial^2_{\Phi_{16}}V_{fin}^{AF(1)}(\Phi)\Big\vert_{\Phi=0}\nonu
&=&6\lambda_{ph}(1,16).
\eea
In the continuum limit it is
\be
\lambda_{ph}(1,16)=\lr-3\lr^2\lm(\delta\cD_2-4\delta\bar\cD_2)
+{1\over 6}\lm\partial^2_{\Phi_1}\partial^2_{\Phi_{16}}
V_{fin}^{AF(1)}(\Phi)\Big\vert_{\Phi=0}
\ee
The finite part of the effective potential, $V^{AF(1)}_{fin}(\Phi)$,
is computed in Appendix \ref{finap}.
The corresponding expressions in the ferro- and the paramagnetic phases are
formally the same.

\section{A low energy effective theory}
Our theory with a single quantum field contains two particles and its 
antiferromagnetic vacuum is
in the ultraviolet regime. So it is not obvious that the evolution of the 
coupling constants for the two particle like excitations obeys the 
renormalization group equations which hold for the usual para- or ferromagnetic theories.
In order to obtain the renormalization group equation for the potential
of the model we introduce the running cut-off, $k$, implemented in each restricted
Brillouin zone in a spherical symmetric manner, 
\be
{\cal D}_\alpha(k)=\biggl\{\biggl(p-{\Lambda\over2}n(\alpha)\biggr)^2\le k^2\biggr\},
\ee
where $\Lambda=2\pi/a$ and the contributions coming from the edges of the toroidal Brillouin 
zones are left out. We approximate the dispersion relation in ${\cal D}_1(k)$ and
${\cal D}_{16}(k)$ with an $O(4)$ invariant parabola and neglect the non-particle like 
excitations. These approximations involve irrelevant operators of the perturbative
continuum limit which should not influence the finite energy behavior. Thus the renormalization 
group equation \cite{senben} for the potential is 
\bea
k\partial_kV^P_k(\Phi_1,\Phi_{2^d})&=&\hf\Omega_dk^d
\ln\biggl\{\biggl[k^2+\tmr(1)+\partial^2_{\Phi_1}V^P_k(\Phi)\biggr]\nonu
&&\biggl[k^2+\tmr(2^d)+\partial^2_{\Phi_{2^d}}V_k^P(\Phi)\biggr]\nonu
&&-\partial_{\Phi_{2^d}}\partial_{\Phi_1}V^P_k(\Phi)\biggr\}\nonu
V^F_k(\Phi)&=&V^P_k(\Phi+\vphi_F),\nonu
V^{AF}_k(\Phi)&=&V^P_k(\Phi+\vphi_{AF}),
\eea
in the leading order of the gradient expansion where $\Omega_d$ stands for the solid
angle in $d$ dimensions. The coefficients of the higher order terms in $k^2$
in the logarithm are kept fixed in our approximation.

Consider now the following renormalizable continuum lagrangian for a scalar and a pseudoscalar
field, $\tilde\phi_+(x)$, $\tilde\phi_-(x)$, respectively, with momentum space cut-off,
\be
L=\hf(\partial_\mu\tilde\phi_+)^2+\hf(\partial_\mu\tilde\phi_-)^2
+V(\tilde\phi_+,\tilde\phi_-),\label{anth}
\ee
whose renormalization group equation in the leading order of the gradient expansion is
\bea
k\partial_kV_k(\tilde\phi_+,\tilde\phi_-)&=&\hf\Omega_dk^d
\ln\biggl\{\bigl[k^2+\partial^2_{\tilde\phi_+}V_k(\tilde\phi)\bigr]
\bigl[k^2+\partial^2_{\tilde\phi_-}V_k(\tilde\phi)\bigr]\nonu
&&-\partial_{\tilde\phi_-}\partial_{\tilde\phi_+}V_k(\tilde\phi)\biggr\}.
\eea
The renormalization group flow of this model agrees with our higher derivative
theory at low energies when the initial condition 
\be
V_{\Lambda}(\tilde\phi_+,\tilde\phi_-)={m^2_{B+}\over2}\tilde\phi_+^2
+{m^2_{B-}\over2}\tilde\phi_-^2
+{\lambda_B\over4}(\tilde\phi_+^4+\tilde\phi_-^4+6\tilde\phi_+^2\tilde\phi_-^2)\label{anpot}
\ee
is chosen. In other words, the model (\ref{anth}), (\ref{anpot}) is equivalent with
(\ref{lagrc}) at low energy when the continuum limit is taken. The correspondence
between the phases is the following,
\bea
P&\Longleftrightarrow<\tilde\phi_+>=0,&<\tilde\phi_->=0,\nonu
F&\Longleftrightarrow<\tilde\phi_+>\not=0,&<\tilde\phi_->=0,\nonu
AF&\Longleftrightarrow<\tilde\phi_+>=0,&<\tilde\phi_->\not=0.
\eea

The conserved momentum of the antiferromagnetic phase is (\ref{consm}) and the
exchange of the momentum $\pi n(16)_\mu/2$ on the lattice with the vacuum corresponds
to the exchange of the scalar and the pseudoscalar particle. Due to the vertex
$\tilde\phi_+^2\tilde\phi_-^2$ in the lagrangian a pseudoscalar particle
can decay into two scalar ones in the antiferromagnetic phase and the parity
is not conserved.

The one-loop scaling laws of our theory agree with a usual two component
$\phi^4$ model up to
irrelevant terms. Thus one may suppose that our theory is not asymptotically free
and consequently becomes trivial in the
continuum limit. In this case when the cut-off can not be eliminated
from the interacting theory the irrelevant terms which were neglected
in the comparison might be important and generate different physical content.

\section{Conclusions}
The one-loop vacuum polarization effects were studied in the para-, ferro- and
$(1,2)$ antiferromagnetic phases of the four dimensional $\phi^4$ model
around the chiral invariant critical point. One can identify two particle like
excitations in each phase. The one-loop divergences were eliminated by an
appropriate fine tuning of the parameters of the bare lagrangian and
the resulting theory was found to be equivalent at low energies with 
a usual renormalizable model made by a scalar and a pseudoscalar field.
In this continuum limit where the length scale of antiferromagnetic vacuum or 
the pseudoscalar staggered particle mode tends to zero the well known problems 
about the unitarity \cite{unit} disappear. 

One should emphasize that even though the cut-off can be removed and the
continuum limit can be taken at the one-loop level the theory can be
defined by relying heavily on the regulator. The renormalized continuum theory
exists only when the regulator is taken into account both at the tree- and
loop-levels in a systematical manner.

The possibility of removing the divergences in the presence of an
apparently non-renormalizable term in the lagrangian is in principle a 
serious threat for the usual strategy of Particle Physics where the
universality is used to limit our investigations to the class of 
renormalizable theories. 
But the result that our model reproduces the infrared structure of a conventional 
renormalizable one is reassuring because it indicates that there is no new
universality class encountered.

The antiferromagnetic phase is certainly different compared to the  
usual $\phi^4$ model with $c_j=0$. But even the para- and the ferromagnetic phases
in our higher derivative model become unusual, as well, in the vicinity of the chiral 
invariant critical point.
This is because the dispersion relation develops in all of these phases a second 
minimum which
can be fine tuned around this critical point in such a manner that another
particle like excitation, the analogue of the rotons of superfluids,
appears. This particle has staggered excitation modes which allow us
to introduce the chiral fields which are exchanged between each other
under the space-time 
inversions. These chiral fields decouple in the chiral symmetrical theory.
By considering field variables only in the sublattice
of the even lattice one can construct models with a single chiral
boson. Such a decoupling of the modes is reminiscent of the fermion
doubling on the lattice and the resulting model with a single chiral
boson is local and satisfies the reflection positivity.

Our computation was made at the one-loop level only. There is no conceptual
problem in extending our work to higher loop orders though the treatment of the
overlapping divergences with unconventional dispersion relation represents
a challenging problem.  
It remains to be seen if the perturbative elimination of the divergences can be achieved 
beyond the one-loop order. If a theory in the antiferromagnetic phase turns out to be 
renormalizable then its vacuum appears homogeneous in physical measurements. It is
the structure of the excitations only which betrays the non-trivial structure of the vacuum
of such a theory. 
There are numerical indications of the continuum limit in the 
antiferromagnetic phase for other models with antiferromagnetic vacuum \cite{lcm}.

One should mention that there are other possible continuum limits in our model
away from the chiral invariant critical region when the mass parameter
is kept at a cut-off independent value. This parameter plays a role analogous
of the $\kappa$-parameter of the Wilson fermions. In fact, the excitations
of the restricted zone ${\cal B}_1$ decouple when $\ml=O(a^0)\not=0$. The only left over
excitations in ${\cal B}_{16}$ become critical along the P-AF transition line.
Thus the approach of the critical line with a fixed mass parameter results
in a theory which
contains a single pseudoscalar particle. In a similar manner certain regions
of the $c_4\not=0$ part of the phase diagram may become critical and
offer a continuum limit. This is because the renormalizability is a rather
straightforward issue when only one particle is left in our model.

Finally we mention the problem of triviality. It is a frustrating experience
that the simplest models such as the $\phi^4$ and QED which are used
in the textbooks to demonstrate the renormalization of Quantum Field Theories
might well be non-renormalizable if they are found to be trivial. In this
case the study of their ultraviolet scaling behaviour serves 
phenomenological interest and a real ultraviolet fixed point can be
achieved by asymptotically free models only. The one-loop ultraviolet
structure of our theory turned out ot be similar to the conventional
$\phi^4$ model. This leads us to assume that our theory is not asymptotically 
free and perhaps trivial because its coupling constant which is marginal
at the tree-level becomes irrelevant due to the one-loop contributions.
This suggests the extension of the investigation of the antiferromagnetic 
vacuum to other, more involved asymptotically free models which may preserve their 
renormalizability and offer a more consistent example of a non-homogeneous
vacuum which actually appears homogeneous in the experiments.

\begin{figure}
\begin{minipage}{5.5cm}
	\epsfxsize=5cm
	\epsfysize=5cm
	\centerline{\epsffile{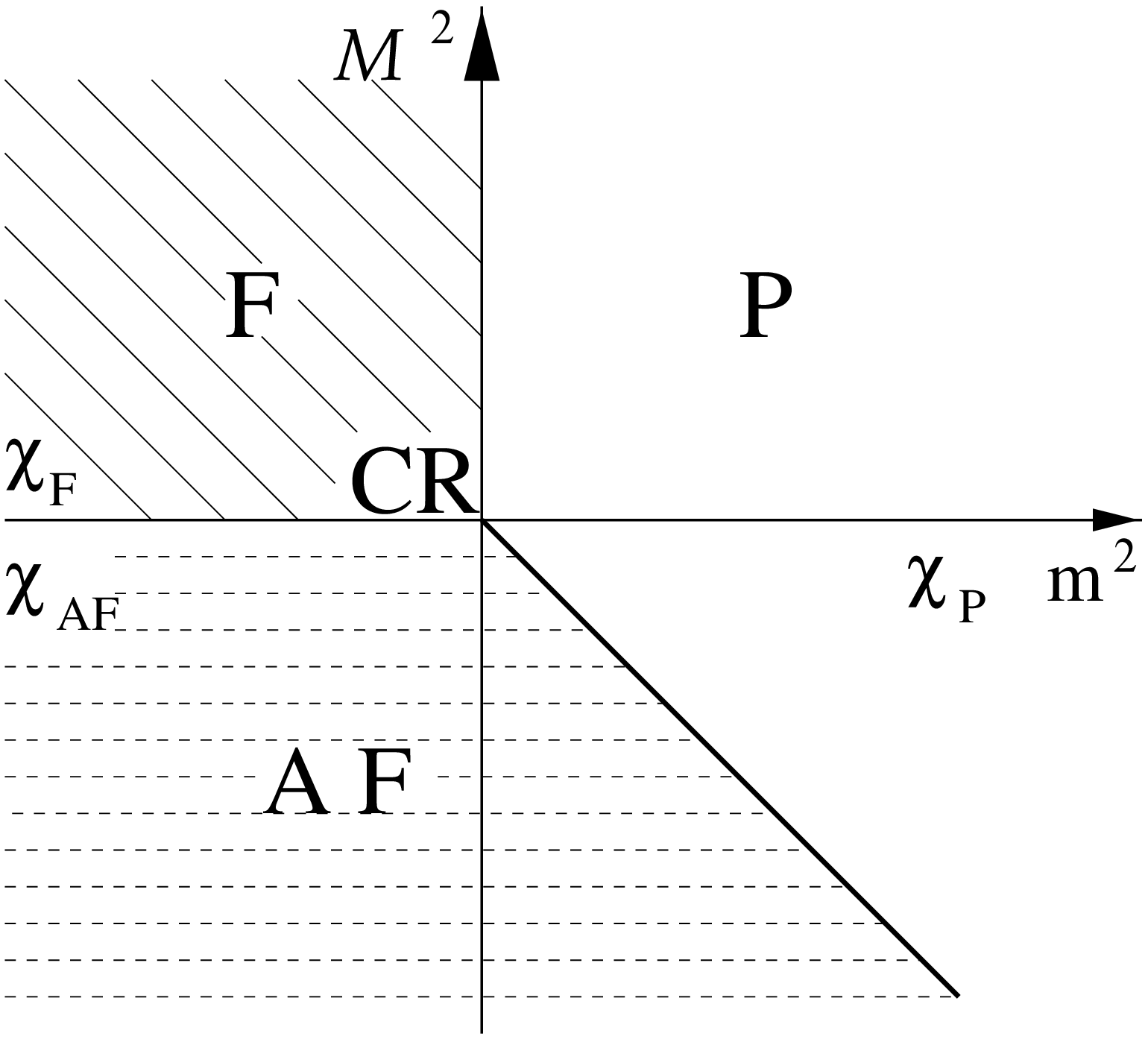}}
\centerline{(a)}
\end{minipage}
\hfill
\begin{minipage}{5.5cm}
	\epsfxsize=5cm
	\epsfysize=5cm
	\centerline{\epsffile{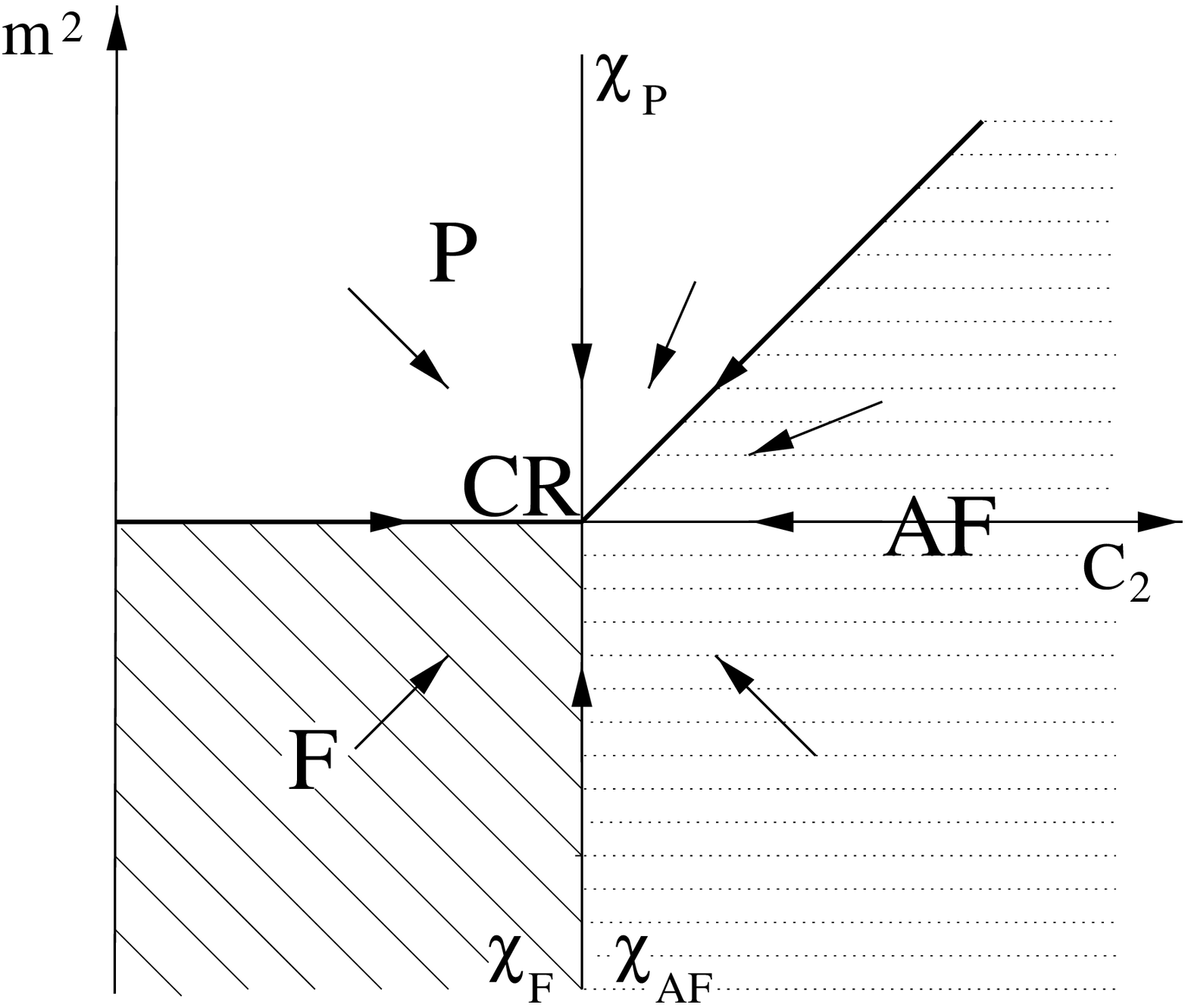}}
\centerline{(b)}
\end{minipage}
\vskip 30pt
\caption{The phase boundary between the paramagnetic (P), ferromagnetic (F) 
and the antiferromagnetic (AF) phase for $c_4=0$. The two particles are degenerate
along the chiral symmetric lines $\chi_P$, $\chi_F$, $\chi_{AF}$.   
(a): The plane $(\ml,\cm)$; (b): The plane $(c_2,\ml)$.
The arrows show the different continuum limits at the critical point CR.}
\end{figure}

\begin{figure}[t]
	\vspace{0.5cm}
	\epsfxsize=8cm
	\epsfysize=2cm
	\centerline{\epsffile{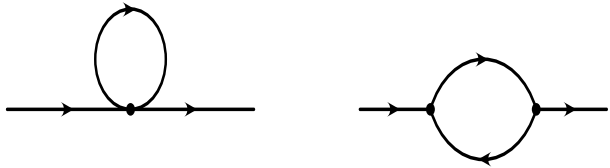}}
\caption{The one-loop self-energy graphs.}
\end{figure}

\begin{table}
\begin{tabular}{|c||c|c|c|c|}
%\begin{tabular}{@{}*{5}{|l||l|l|l|l}}
\hline    
phase&$\tml(1)$&$\tml(16)$&$Z(1)$&$Z(16)$\cr 
\hline    
P &$\ml$       &$\ml+\cm$   &1&$-1+32c_2$\cr
F &$-2\ml$     &$-2\ml+\cm$ &1&$-1+32c_2$\cr
AF&$-2\ml-3\cm$&$-2\ml-2\cm$&1&$-1+32c_2$\cr
\hline
\end{tabular}
\vskip 30pt
\caption{The parameters of the propagator for ${\cal B}_1$ and ${\cal B}_{16}$.}
\end{table}

\begin{table}
\begin{tabular}{|c||c|c|c|c|c|}
%\begin{tabular}{@{}*{6}{|l||l|l|l|l|l}}
\hline
phase&$\kappa$&$\sigma$&$\mr$&$\tmr(16)$&$\tmr(1)-\tmr(16)$\cr
\hline    
$P_F$      &0   &$-1$  &$>0$      &$\mr+16\mu^2(\kappa)$  &$-16\mu^2(\kappa)$\cr
$\chi_P$   &$>0$&$\pm1$&$>0$      &$\mr$                  &0\cr
$P_{AF}$   &0   &$+1$  &$>16\mu^2$&$\mr-16\mu^2(\kappa)$  &$16\mu^2(\kappa)$\cr
$P-AF$     &0   &$+1$  &$=16\mu^2$&0                      &$16\mu^2(\kappa)$\cr
$AF$       &0   &$+1$  &$<16\mu^2$&$-2\mr+32\mu^2(\kappa)$&$16\mu^2(\kappa)$\cr
$\chi_{AF}$&$>0$&$+1$  &$<0$      &$-2\mr$                &0\cr
$\chi_F$   &$>0$&$-1$  &$<0$      &$-2\mr$                &0\cr
$F$        &0   &$-1$  &$<0$      &$-2\mr+16\mu^2(\kappa)$&$-16\mu^2(\kappa)$\cr
\hline
\end{tabular}
\vskip 30pt
\caption{The different ways of approaching the critical point of Fig.1b. $P_F$ 
and $P_{AF}$ are the regions in the paramagnetic phase respectively on the left and on 
the right side of the chiral line $\chi_P$.}
\end{table}

\acknowledgments
We thank Jan Stern for interesting discussions.

\appendix
\section{Computation of the effective potential}\label{aefint}
\underline{The tree-level:}The tree-level effective potential for the background field 
\be
\Phi\br=\Phi_1\gamma^1+\Phi_{2^d}\gamma^{2^d},
\ee
in the antiferromagnetic phase is the sum of the renormalized potential and the counterterms,
\bea
V^{AF(0)}(\Phi)&=&\hf\biggl(\ml+\dml+3(\lr+\dl)\vphi_{2^d}^2\biggr)\Phi^2_1\nonu
&&+\hf\biggl(P^2(2^d)\cK(-P^2(2^d))
+\ml+\dml+3(\lr+\dl)\vphi^2_{2^d}\biggr)\Phi_{2^d}^2\nonu
&&+(\dml+\dl\vphi^2_{2^d})\vphi_{2^d}\Phi_{2^d}+
\vphi_{2^d}(\lr+\dl)(\Phi_{2^d}^3+3\Phi_1^2\Phi_{2^d})\nonu
&&+{\lr+\dl\over4}(\Phi_1^4+\Phi_{2^d}^4+6\Phi_1^2\Phi_{2^d}^2),
\eea
what can be written up to a constant as 
\bea 
V^{AF(0)}(\Phi)&=&\hf\biggl(P^2(1)\cK(-P^2(1))+\ml+\dml\biggr)\Phi^2_1\nonu
&&+\hf\biggl(P^2(2^d)\cK(-P^2(2^d))+\ml+\dml\biggr)(\Phi_{2^d}+\vphi_{2^d})^2\nonu
&&+{\lr+\dl\over4}[\Phi_1^4+(\Phi_{2^d}+\phi_{2^d})^4]
+{3\over2}(\lr+\dl)\Phi_1^2(\Phi_{2^d}+\phi_{2^d})^2.\nonu\label{varba} 
\eea

\underline{The one-loop level:} The next step is to obtain the one-loop contribution,
\bea
V^{AF(1)}_{eff}(\Phi)=
&&\hf\int_{p\le\pi/2}{d^dp\over(2\pi)^d}\tr\ln\biggl[(P+p)^2\cK(-(P+p)^2)\nonu
&&+\tml+6\lr(\Phi_1+\gamma^{2^d}\Phi_{2^d})\vphi_{2^d}\gamma^{2^d}\nonu
&&+3\lr(\Phi_1+\gamma^{2^d}\Phi_{2^d})^2\biggr].
\eea
Since $(\gamma^{2^d})^2=\gamma^1=1$ we can write this integral as
\be
\int_{p\le\pi/2}{d^dp\over(2\pi)^d}\ln\det[A(p)+B\gamma^{2^d}],
\ee
where
\bea
A(p)+B\gamma^{2^d}&=&
\pmatrix{
A_1(p)&&0\cdots&0&0\cr
0&&A_2(p)\cdots&0&0\cr
\vdots&\vdots&\ddots&\vdots&\vdots\cr
0&0&\cdots&A_{2^d-1}(p)&0\cr
0&0&\cdots&0&A_{2^d}(p)\cr}\nonumber\\
&+&\pmatrix{
0&0&\cdots&0&B\cr
0&0&\cdots&B&0\cr
\vdots&\vdots&\adots&\vdots&\vdots\cr
0&B&\cdots&0&0\cr
B&0&\cdots&0&0\cr}
\eea
and
\bea
A_\alpha(p)&=&G^{-1}_\alpha(p)+C,\nonu
B&=&6\lr\Phi_1(\vphi_{2^d}+\Phi_{2^d}),\nonu
C&=&3\lr(2\vphi_{2^d}\Phi_{2^d}+\Phi_{2^d}^2+\Phi^2_1).\label{reszl}
\eea
The determinant in question is
\be
\det[A(p)+B\gamma^{2^d}]=\prod_{\alpha=1}^{2^{d-1}}
(A_\alpha A_{\bar\alpha}-B^2).
\ee
In this manner we obtain
\bea
V^{AF(1)}_{eff}(\Phi)&=&\hf\int_{p\le\pi/2}{d^dp\over(2\pi)^d}
\ln\det[A(p)+B\gamma^{2^d}]\nonu
&=&\hf\int_{p\le\pi/2}{d^dp\over(2\pi)^d}
\sum_{\alpha=1}^{2^{d-1}}\ln[A_\alpha(p)A_{\bar\alpha}(p)-B^2].\label{effre}
\eea
In order to isolate the UV divergences it is advantageous to write
\bea
V^{AF(1)}_{eff}(\Phi)&=&\hf\int_{p\le\pi/2}{d^4p\over(2\pi)^4}
\sum_{\alpha=1}^{2^{d-1}}\ln\bigl[(G^{-1}_\alpha+C)(G^{-1}_{\bar\alpha}+C)
-B^2\bigr]\nonu
&=&\int_{p\le\pi/2}{d^4p\over(2\pi)^4} 
\sum_{\alpha=1}^{2^{d-1}}\ln\bigl[1+CG_\alpha +CG_{\bar\alpha}
+(C^2-B^2)G_\alpha G_{\bar\alpha}\bigr] \nonumber\\
&+&\int_{p\le\pi/2}{d^4p\over(2\pi)^4}\sum_{\alpha=1}^{2^{d-1}}
\ln\bigl[G^{-1}_\alpha G^{-1}_{\bar\alpha}\bigr].\label{pot}
\eea
The detailed form of (\ref{effre}) reads as
\bea
V^{AF(1)}_{eff}(\Phi)&=&\hf\int_{p\le\pi/2}{d^dp\over(2\pi)^d}
\sum_{\alpha=1}^{2^{d-1}}\ln\biggl\{\biggl[
(P(\alpha)+p)^2\cK(-(P(\alpha)+p)^2)\nonu
&&+\tilde m^{2AF}_{LR}+3\lr(2\vphi_{2^d}\Phi_{2^d}+\Phi_{2^d}^2+\Phi^2_1)\biggr]\times\nonu
&&\biggl[(P(\bar\alpha)+p)^2\cK(-(P(\bar\alpha)+p)^2)\nonu
&&+\tilde m^{2AF}_{LR}+3\lr(2\vphi_{2^d}\Phi_{2^d}+\Phi_{2^d}^2+\Phi^2_1)\biggr]\nonu
&&-36\lambda^2_R\Phi_1^2(\vphi_{2^d}+\Phi_{2^d})^2\biggr\}.\label{effpoti}
\eea
The corresponding expressions for the ferromagnetic phase are obtained by the
exchange $1\leftrightarrow 2^d$, 
\bea
V^{F(1)}_{eff}(\Phi)&=&\hf\int_{p\le\pi/2}{d^dp\over(2\pi)^d}
\sum_{\alpha=1}^{2^{d-1}}\ln\biggl\{\biggl[
(P(\alpha)+p)^2\cK(-(P(\alpha)+p)^2)\nonu
&&+\tilde m^{2F}_{LR}+3\lr(2\vphi_1\Phi_1+\Phi_{2^d}^2+\Phi^2_1)\biggr]\times\nonu
&&\biggl[(P(\bar\alpha)+p)^2\cK(-(P(\bar\alpha)+p)^2)\nonu
&&+\tilde m^{2F}_{LR}+3\lr(2\vphi_1\Phi_1+\Phi_{2^d}^2+\Phi^2_1)\biggr]\nonu
&&-36\lambda^2_R(\vphi_1+\Phi_1)^2\Phi_{2^d}^2\biggr\}.\label{effpotf}
\eea
In the paramagnetic phase one sets $\vphi_\alpha=0$,
\bea
V^{P(1)}_{eff}(\Phi)&=&\hf\int_{p\le\pi/2}{d^dp\over(2\pi)^d}
\sum_{\alpha=1}^{2^{d-1}}\ln\biggl\{\biggl[
(P(\alpha)+p)^2\cK(-(P(\alpha)+p)^2)\nonu
&&+\ml+3\lr(\Phi_{2^d}^2+\Phi^2_1)\biggr]\times\nonu
&&\biggl[(P(\bar\alpha)+p)^2\cK(-(P(\bar\alpha)+p)^2)\nonu
&&+\ml+3\lr(\Phi_{2^d}^2+\Phi^2_1)\biggr]\nonu
&&-36\lambda^2_R\Phi_1^2\Phi_{2^d}^2\biggr\}.\label{effpotp}
\eea

\section{Expansion around the symmetrical theory}\label{mudep}
We study in this Appendix the dependence of the loop integrals on the scale 
parameter $\mu$ what controls the strength of the explicit breaking of the chiral
symmetry. For this end we need the continuum limit for the integrals of the type
\be
\lm I_n=\lm\int_{p\le{\pi\over2a}}d^4p G_\alpha^n(p).
\ee
Close to the symmetrical point the propagator has two maxima, namely in the 
regions $\alpha=1, 16$. In case of a convergent integral one expects that 
most of the contribution comes from the regions around these maxima. 
But due to the presence of divergencies the problem has to be considered more carefully.

\underline{The Brillouin zone ${\cal B}_1:$} The $\mu$-dependence shows up in the
antiferromagnetic phase only. One expands around $p=0$,
\be
G_1^{-n}(p)=\biggl(p^2+\tmr(1)+ba^2p^4+ca^4p^6+O(a^6p^8)\biggr)^n
\ee
where 
\bea
b&=&b_0+b_1(\mu a)^{2+\kappa}\nonu
c&=&c_0+c_1(\mu a)^{2+\kappa}
\eea
are dimensionless functions. One finds
\bea
\int d^4pG_1^n(p)&=&\int d^4p{1\over(p^2+\tmr(1))^n}\nonu
&&\times\biggl[1-n{ba^2p^4+ca^4p^6+O(a^6p^8)\over p^2+\tmr(1)}\\
&&+{n(n+1)\over2}\biggl({ba^2p^4+ca^4p^6+O(a^6p^8)\over
p^2+\tmr(1)}\biggr)^2+\cdots\biggr].\nonumber
\eea
The corrections to the usual first term are of the form 
\be
\int_{-{\pi\over2a}}^{{\pi\over2a}}d^4p{a^kp^\ell\over(p^2+\tmr(1))^{n+m}}
\ee
where $\ell-k-2m=0$ in order to keep the dimension of each contribution.
The one-loop integral is finite for $4-2(n+m)+\ell-k<0$ i.e. for $n\ge3$.
Since $k\ge0$ the finiteness implies vanishing corrections and $\mu$-independence,
\be
\lm\int_{-{\pi\over2a}}^{{\pi\over2a}}d^4pG_1^n(p)=\int_{-\infty}^\infty d^4p(p^2+\tmr(1))^{-n},
\ee
for $n\ge3$. The non-vanishing corrections arise for $n=1$ and 2.

The mass renormalization can be carried out without difficulties for 
$\mu\not=0$. In order to follow the renormalization of the coupling constant 
we need the $\mu$-dependence for $n=2$,
\bea
\lm\int_{-{\pi\over2a}}^{{\pi\over2a}}{d^4p\over(2\pi)^4}G_1^2(p)
&=&\int_{-\infty}^\infty{d^4p\over(2\pi)^4}{1\over(p^2+\tmr(1))^2}+I_{fin}^\star\nonu
&=&{1\over16\pi^2}\biggl(\ln{\Lambda^2\over\tmr(1)}-1\biggr)
+\tilde I_{fin}^\star\label{kvnk}
\eea
where $\Lambda=2\pi/a$ and $I_{fin}$ is a finite function of $\mu$.
Since $I_{fin}$ depends on $\mu$ through the combination $(\mu a)^{2+\kappa}$
and the UV divergence is logarithmic only the finite part of (\ref{kvnk}) 
becomes $\mu$-independent in the continuum limit.

\underline{The  Brillouin zone ${\cal B}_{16}:$} The mass is $\mu$-dependent
in each phase and we write $\tmr(16)=\tmr(1)+\tmrp$ where
\be
\tmrp=-16\sigma\mu^2(a\mu)^\kappa.
\ee
The expansion is made around $p=P(16)$,
\be
G_{16}^{-n}(p)=\biggl(p^2+\tmr(1)+\tmrp+ba^2p^4+ca^4p^6
+O(a^6p^8)\biggr)^n,
\ee
\bea
\label{ntepsk} 
\int d^4pG_{16}^n(p)&=&\int d^4p{1\over(p^2+\tmr(1))^n}\nonu
&&\times\biggl[
1-n{\tmrp+ba^2p^4+ca^4p^6+O(a^6p^8)\over p^2+\tmr(1)}\\
&&+{n(n+1)\over2}\biggl({\tmrp+ba^2p^4+ca^4p^6+O(a^6p^8)\over
p^2+\tmr(1)}\biggr)^2+\cdots\biggr].\nonumber
\eea 
The repetition of the argument followed in the previous case yields to the
$\mu$-independent finite result,
\be
\lm\int_{-{\pi\over2a}}^{{\pi\over2a}}d^4pG_{16}^n(p)
=\int_{-\infty}^\infty{d^4p\over(p^2+\tmr(1)+\tmrp)^n},
\ee
for $n\ge3$. For $n=2$ one finds
\bea
\lm\int_{-{\pi\over2a}}^{{\pi\over2a}}{d^4p\over(2\pi)^4}G_{16}^2(p)
&=&\int_{-\infty}^\infty{d^4p\over(2\pi)^4}{1\over(p^2+\tmr(1)+\tmrp)^2}
+I_{fin}^\star\nonu
&=&{1\over16\pi^2}\biggl(\ln{\Lambda^2\over\tmr(1)+\tmrp}-1\biggr)+I_{fin}^\star\\
&=&{1\over16\pi^2}\biggl(\ln{\Lambda^2\over\tmr(1)}-\ln{\tmr(16)\over\tmr(1)}-1\biggr)
+I_{fin}^\star.\nonumber
\eea
There is a similar result for the mixed product,
\bea
\lm\int_{-{\pi\over2a}}^{{\pi\over2a}}{d^4p\over(2\pi)^4}G_1(p)G_{16}(p)
&=&\int_{-\infty}^\infty{d^4p\over(2\pi)^4}{1\over(p^2+\tmr(1))(p^2+\tmr(1)+\tmrp)}\nonu
&&+I_{fin}^\star\\
&=&{1\over16\pi^2}\biggl(\ln{\Lambda^2\over\tmr(1)}-{\tmr(16)\over\tmrp}
\ln{\tmr(16)\over\tmr(1)}\biggr)\nonu
&&+I_{fin}^\star.\nonumber
\eea

\underline{The regions ${\cal B}_\alpha$, $\alpha=2,\cdots,15$:} 
We can find a real number, $\gamma$, such that
\be
G_\alpha(p)\le\gamma a^2,\label{gama}
\ee
or
\be
\int_{{\pi\over2a}}^{{\pi\over2a}}{d^4p}G_\alpha^n(p)\le\gamma^na^{2n}\int_{{\pi\over2a}}^{{\pi\over2a}}d^4p.
\ee
This integral is vanishing in the continuum limit for $n\ge3$. For $n=2$ 
one has
\bea
G_\alpha(p)G_\beta(p)&=&G^{\star}_\alpha(p) G^{\star}_\alpha(p)
\biggl[1+\sigma\mu^\kappa a^{\kappa-2}\biggl(
\sin^4(p+P(\alpha))G^{\star}_\beta(p)\nonu
&&+\sin^4(p+P(\beta))G^{\star}_\alpha(p\biggr)+...\biggr].
\eea
From
\be
\mu^na^{n-2}\sin^4(p+P(\alpha))\le\mu a^{n-2}
\ee
and
\be
G^\star_\beta(p)\le \gamma a^2
\ee
one obtains
\be
\mu^na^{n-2}\sin^4(p+P(\alpha))G^{3\star}_\beta(p)\le 
\mu\gamma^3a^{n+4}
\ee
which yields the equation
\bea
\lm\int_{{\pi\over2a}}^{{\pi\over2a}}d^4pG_\alpha(p)G_\beta(p)=
\int_{{\pi\over2a}}^{{\pi\over2a}}d^4pG^\star_\alpha(p)G^\star_\beta(p) 
\eea
in the continuum limit.

So one finds the divergent part
\bea
\cD_2&=&\cD_2^\star+\delta\cD_2,\nonu
\bar\cD_2&=&\bar\cD^\star_2+\delta\bar\cD_2,
\eea
with $\cD_2^\star=2\bar\cD_2^\star$ and
\bea
\delta\cD_2&=&-{1\over16\pi^2}\ln{\tmr(16)\over\tmr(1)}\nonu
\delta\bar\cD_2&=&{1\over16\pi^2}\biggl(1-{\tmr(1)+\tmrp
\over\tmrp}\ln{\tmr(16)\over\tmr(1)}\biggr).
\eea

\section{The finite part of the effective potential}\label{finap}
To compute the finite part of the effective potential we start from 
equation (\ref{pot}), separate the finite contributions and cancel the 
divergences by the counterterms. Actually the finite part of the potential 
in the continuum limit depends only the terms containing $G_1$ et $G_{16}$.

\underline{Non-degenerate masses, $(\kappa=0)$:} We seek
\bea
V^{AF(1)}_{fin}(\Phi)&=&\hf\int_{p\le\Lambda}{d^4p\over(2\pi)^4}
\ln\bigl[(p^2+\tmr(1)+C)(p^2+\tmr(16)+C)-B^2\bigr]\nonu
&&-{C\over2}\int_{p\le\Lambda}{d^4p\over(2\pi)^4}\biggl({1\over p^2+\tmr(1)} 
+{1\over p^2+\tmr(16)}\biggr)\nonu
&&+{C^2\over4}\int_{p\le\Lambda}{d^4p\over(2\pi)^4}\biggl(
{1\over(p^2+\tmr(1))^2}+{1\over(p^2+\tmr(16))^2}\biggr)\nonu
&&+{B^2\over2}\int_{p\le\Lambda}{d^4p\over(2\pi)^4}
{1\over(p^2+\tmr(1))(p^2+\tmr(16))},\label{ndmr}
\eea
which gives after integration
\bea
V^{AF(1)}_{fin}(\Phi)&=&{1\over128\pi^2}
\biggl\{(a^2_{nd}-b^2_{nd})\ln{\tmr(16)\over\tmr(1)}\nonu
&&+(a^2_{nd}+b^2_{nd}+d_{nd})\ln{2a_{nd}b_{nd}-d_{nd}\over2\tmr(1)\tmr(16)}\nonu
&&+d_{nd}{\tmr(16)+\tmr(1)\over\tmr(1)-\tmr(16)}\ln{\tmr(16)\over\tmr(1)}
-2c_{nd}(\tmr(16)+\tmr(1))\nonu
&&-6c^2_{nd}-d_{nd}-(a_{nd}+b_{nd})\sqrt{(a_{nd}-b_{nd})^2+2d_{nd}}\nonu
&&\times\ln{a_{nd}+b_{nd}-\sqrt{(a_{nd}-b_{nd})^2+2d_{nd}}\over
a_{nd}+b_{nd}-\sqrt{(a_{nd}-b_{nd})^2+2d_{nd}}}\biggr\},
\eea
with
\bea
a_{nd}&=&\tmr(1)+3\lr(2\vphi_{16}\Phi_{16}+\Phi_{16}^2+\Phi_1^2),\nonu
b_{nd}&=&\tmr(16)+3\lr(2\vphi_{16}\Phi_{16}+\Phi_{16}^2+\Phi_1^2),\nonu
c_{nd}&=&3\lr(2\vphi_{16}\Phi_{16}+\Phi_{16}^2+\Phi_1^2),\nonu
d_{nd}&=&72\lr^2\Phi_1^2(\vphi_{16}+\Phi_{16})^2,\nonu
\vphi_{16}^2&=&{1\over\lr}(-\mr+16\mu^2).
\eea
We find 
\bea
\partial^2_{\Phi_1}V^{AF(1)}_{fin}(\Phi)\Big\vert_{\Phi=0}
&=&\partial^2_{\Phi_{16}}V^{AF(1)}_{fin}(\Phi)\Big\vert_{\Phi=0}=0,\nonu
\partial^4_{\Phi_1}V^{AF(1)}_{fin}(\Phi)\Big\vert_{\Phi=0}
&=&{\lr^3\vphi_{16}^2\over128\pi^2}
\biggl\{{124416\lr\vphi_{16}^2\over(\tmr(1)-\tmr(16))^2}\nonu
&&+{10368\over\tmr(1)-\tmr(16)}\ln{\tmr(16)\over\tmr(1)}\nonu
&&+62208\lr\vphi_{16}^2{\tmr(16)+\tmr(1)\over(\tmr(1)-\tmr(16))^3}
\ln{\tmr(16)\over\tmr(1)}\biggr\},\nonu
\partial^4_{\Phi_{16}}V^{AF(1)}_{fin}(\Phi)\Big\vert_{\Phi=0}
&=&{5184\lr^3\vphi_{16}^2\over128\pi^2}\biggl({1\over\tmr(1)+\tmr(16)}
-{\lr\vphi_{16}^2\over\tilde m^4_R(1)+\tilde m^4_R(16)}\biggr),\nonu
\partial^4_{\Phi_1}\partial^2_{\Phi_{16}}V^{AF(1)}_{fin}\Big\vert_{\Phi=0}
&=&{\lr^3\vphi_{16}^2\over128\pi^2}\biggl\{{864\over\tmr(1)+\tmr(16)}
-{1728\lr\vphi_{16}^2\over\tmr(1)\tmr(16)}\nonu
&&-{576\lr\vphi_{16}^2\over(\tmr(1)-\tmr(16))^2}\\
&&+3456(\tmr(1)-\tmr(16))\ln{\tmr(1)\over\tmr(16)}\biggr\}.\nonumber
\eea

\underline{Degenerate masses, $(\kappa>0)$:} If the two masses are degenerate
then the effective potential is
\bea
V_{fin}^{AF(1)}(\Phi)&=&{1\over64\pi^2}\biggl[(a^2_d+b^2_d)\ln{a^2_d-b^2_d\over\tilde m_R^4}
-3b^2_d-2\tmr c_d-3c^2_d\nonu
&&-2a_db_d\ln{a_d-b_d\over a_d+b_d}\biggr],
\eea
with 
\bea
\tmr&=&-2\mr\nonu
a_d&=&\tmr+3\lr(2\vphi_{16}\Phi_{16}+\Phi_{16}^2+\Phi_1^2),\nonu
b_d&=&6\lr\Phi_1(\vphi_{16}+\Phi_{16}),\nonu
c_d&=&3\lr(2\vphi_{16}\Phi_{16}+\Phi_{16}^2+\Phi_1^2),\nonu
\vphi_{16}^2&=&-{\mr\over\lr}.
\eea
In this case
\be
\partial^4_{\Phi_1}V^{AF(1)}_{fin}\Big\vert_{\Phi=0}
=\partial^4_{\Phi_{16}}V^{AF(1)}_{fin}\Big\vert_{\Phi=0} 
=\partial^2_{\Phi_1}\partial^2_{\Phi_{16}}V^{AF(1)}_{fin}
\Big\vert_{\Phi=0}=648\lr^2.
\ee

The effective potential and its derivatives of the ferromagnetic phase
can be obtained from the corresponding formulae of the antiferromagnetic phase
by exchanging the index $1\longleftrightarrow16$. In the paramagnetic
phase one finds 
\be
\partial^2_{\Phi_1}V^{P(1)}_{fin}\Big\vert_{\Phi=0}
=\partial^2_{\Phi_{16}}V^{P(1)}_{fin}\Big\vert_{\Phi=0}=0.
\ee
Observe that the particle of the zone ${\cal B}_{16}$ remains massless
along the P-AF transition line.

\end{document}